# About Lie group of the electron magnetohydrodynamics with effect of electron inertia and finite conducting.

E. Avdeeva
*Department of the Theoretical Physics,*
*Kursk State Pedagogical University, Radischeva st. 33, Kursk, Russia.*

The Lie group of symmetry of the one-dimensional differential equation of electron magnetohydrodynamics with effect of electron inertia and finite conducting is found. It's indicated that this group is group of shift of independent variables and non single-valuety of a potential. Therefore thee no automodel solution except for running vawes.

Ellectron magnetohydrodynamics (EMH) is a modern branch of plasma physics. It's the multicomponent magnetohydrodynamics based on quasi-neutrality, the Hall effect and negligible ion motion. EMH has many applications to problems of pulsed plasmas at al. Remarkable effect of EMH is transformed skin problem. The exact solution of this problem is found as chock vawe in virtue of reduced EMH equation without electron inertia to the Burgers equation.[1].

It's well known that one can do some inference about a sulution of non-linear diferential equation knowing it's Lie group. The agregation of Lie groups for the Burgers equation [2] and Lie-Backlund groups for EMH equation with infinite conducting and inertia of electrons [3] are known. But the searcing group for EMH equation with both electron inertia and finite conducting is actual problem. It's the problem of this work.

One-dimension equation for magnetic induction is:

$$\frac{\partial}{\partial t}\left[B-\left(\frac{c}{\omega_{pe}}\right)^2 \frac{\partial^2 B}{\partial z^2}\right] + k\frac{\partial B}{\partial z}\left[B-\left(\frac{c}{\omega_{pe}}\right)^2 \frac{\partial^2 B}{\partial z^2}\right] = D\frac{\partial^2 B}{\partial z^2} \qquad (1)$$

where $\omega_{pe}$ - electron plasma frequensy, D – coefficient of diffusion of magnetic field. It's comfortable for searhing group to write (1) as equation for potential u ($B = u_z$):

$$u_t - (c/\omega_{pe})^2 u_{zzt} + k/2 u_z^2 - k/2(c/\omega_{pe})^2 u_{zz}^2 = D u_{zz} . \qquad (2)$$

The standard technique of group analysis is written in [2]. Applying this technique we have the infinitesimal criterion of invariance for coefficients $\xi$, $\tau$, $\varphi$:

$$\varphi^t - (c/\omega_{pe})^2 \varphi^{zzt} + k\varphi^z u_z - k(c/\omega_{pe})^2 \varphi^{zz} u_{zz} = D\varphi^{zz},$$

where
$$\varphi^z = D_z\varphi - u_z D_z\xi - u_t D_z\tau\ ;\quad \varphi^t = D_t\varphi - u_z D_t\xi - u_t D_t\tau\ ;$$

$$\varphi^{zz} = D_z^2\varphi - u_z D_z^2\xi - u_t D_z^2\tau - 2u_{zz}D_z\xi - 2u_{zt}D_z\tau\ ;$$

$$\varphi^{zzt} = D_z^2 D_t\varphi - u_t D_z^2 D_t\tau - 2u_{zt}D_z D_t\tau - u_{zzt}D_t\tau - u_z D_z^2 D_t\xi - 2u_{zz}D_z D_t\xi - u_{zzz}D_t\xi -$$

$$u_{tt}D_z^2\tau - u_{zt}D_z^2\xi - 2u_{ztt}D_z\tau - 2u_{zzt}D_z\xi.$$

The defining system of equation is:

$$\begin{aligned}
&\xi_u = \xi_t = \tau_u = \tau_z = 0 &&(c/\omega_{pe})^2\varphi_{uzz} + \tau_t - 4\xi_z + \varphi_u = 0\\
&\varphi_{uu} = 0 &&(c/\omega_{pe})^2\varphi_{uzz} + \tau_t - 2\xi_z + \varphi_u = 0 \qquad (3)\\
&2\varphi_{uz} - \xi_{zz} = 0 &&(c/\omega_{pe})^2\varphi_{ut} + D\tau_t = 0\\
&(c/\omega_{pe})^2\varphi_{uzz} - 2\xi_z = 0 &&\varphi_t - (c/\omega_{pe})^2\varphi_{zzt} - D\varphi_{zz} = 0
\end{aligned}$$

The solution of (3) is: $\xi = C_1$, $\tau = C_2$, $\varphi = C_3$, $C_{1,2,3}$ = const.
The vector fields are: $\mathbf{v}_1 = \partial_t$, $\mathbf{v}_2 = \partial_z$, $\mathbf{v}_3 = \partial_u$.
The groups generated those fields are:

$$G_1:(z+\varepsilon, t, u),\ G_2:(z, t+\varepsilon, u),\ G_3:(z, t, u+\varepsilon),\ \varepsilon \in R\ . \qquad (4)$$

The $G_1$ and $G_2$ are groups of shift of independent variables. The $G_3$ is connected with non single-valuety of a potential.

It's visually that group of symmetry of (2) is less than group of symmetry of Burgers equation. It don't admit any linearizing substitution and there is one useful solution as corollary of the invariant of the group. It's a running wave $B = B(z - ct)$. This solution sees in [1].

From the result about Lie group (4) it will be observed that there are no other simplifications.

**Acknowledgments:** Author would like to thank her scientific supervisor E.B. Postnikov.